\documentclass[journal=jacsat,manuscript=article]{achemso}

\usepackage[version=3]{mhchem} 

\usepackage{color, soul}
\usepackage{amsmath}
\usepackage{mathtools}

\DeclareUnicodeCharacter{2212}{+}
\DeclareUnicodeCharacter{0327}{+}

\author{Purbasha Sharangi}
\author{Esita Pandey}
\author{Shaktiranjan Mohanty}
\affiliation[Unknown University]
{Laboratory for Nanomagnetism and Magnetic Materials (LNMM), School of Physical Sciences, National Institute of Science Education and Research (NISER), HBNI, P.O.- Bhimpur Padanpur, Via Jatni, 752050, India}
\author{Sagarika Nayak}
\author{Subhankar Bedanta}
\affiliation[Unknown University]
{Laboratory for Nanomagnetism and Magnetic Materials (LNMM), School of Physical Sciences, National Institute of Science Education and Research (NISER), HBNI, P.O.- Bhimpur Padanpur, Via Jatni, 752050, India}
\alsoaffiliation[Second University]
{Center for Interdisciplinary Sciences (CIS), National Institute of Science Education and Research (NISER), HBNI, Jatni, 752050 India}
\email{sbedanta@niser.ac.in}
\title[An \textsf{achemso} demo]
  {Spinterface Induced Modification in Magnetic Properties in  $Co_{40}Fe_{40}B_{20}$/Fullerene Bilayers}


\begin{document}



\begin{abstract}
  Organic semiconductor/ferromagnetic bilayer thin films can exhibit novel properties due to the formation of spinterface at the interface. Buckminsterfullerene (C$_{60}$) has been shown to exhibit ferromagnetism at the interface when it is placed next to a ferromagnet (FM) such as Fe or Co. Formation of spinterface occurs due to the orbital hybridization and spin polarized charge transfer at the interface. In this work, we have demonstrated that one can enhance the magnetic anisotropy of the low Gilbert damping alloy CoFeB thin film by introducing a $C_{60}$ layer. We have shown that anisotropy increases by increasing the thickness of C$_{60}$ which might be a result of the formation of spinterface. However, the magnetic domain structure remains same in the bilayer samples as compared to the reference CoFeB film.
\end{abstract}

\section{Introduction}
Organic spintronics has drawn immense research interest in the last few decades due to its applications in spin valve, magnetic tunnel junctions etc \cite{naber2007organic,stamps20142014,kuch2016controlling}. In organic spintronics, organic semiconductors (OSCs) (e.g. C$_{60}$, Alq$_{3}$, ruberene etc.) are used to transport or control spin polarized signals \cite{dediu2009spin,atodiresei2010design,barraud2010unravelling,wang2009organic,sun2018progress}. The main advantage of OSCs are their low production cost, light weight, flexible and chemically interactive nature. Usually the spin orbit coupling is small in organic materials (e.g. C$_{60}$) as they consist of low Z (atomic number) materials (in particular carbon (C)). Moreover, the zero hyperfine interaction in C$_{60}$ results in a longer spin relaxation time \cite{moorsom2014spin,tran2011hybridization,tran2013magnetic,djeghloul2016high,gobbi2011room,zhang2013observation,nguyen2013spin}. As a consequence, spin of a carrier weakly interacts in organic environment and spin information is maintained for a long time. There are several reports on organic spin valves, organic light emitting diodes (OLED) using C$_{60}$ as a spacer layer \cite{moorsom2014spin,tran2011hybridization,tran2013magnetic,djeghloul2016high,gobbi2011room,zhang2013observation,nguyen2013spin,liu2018studies}. It has been shown that C$_{60}$ ($\sim$ 2 nm) can be magnetized when it is placed next to a ferromagnetic (FM) layer \cite{moorsom2014spin,mallik2018effect,mallik2019tuning,mallik2019enhanced}. $d-p$ hybridization at the interface of FM/C$_{60}$ modifies the density of states (DOS) and exhibits room temperature ferromagnetism. Such kind of interface is known as spinterface \cite{sanvito2010molecular}. It has been shown that the fundamental magnetic properties like magnetic moment, domain structure and magnetic anisotropy can be tuned by depositing C$_{60}$ on top of a Fe, Co or Fe$_{4}$N layer \cite{mallik2018effect,mallik2019tuning,mallik2019enhanced,han2019spin}. Using first-principles calculations Han $et$ $al.$ have shown that magnetic anisotropy energy (MAE) of Fe$_{4}$N system is changed from out-of-plane to in-plane after inserting a C$_{60}$ layer \cite{han2019spin}. Their study indicates a strong $d-p$ hybridization between Fe and C atoms, which modifies the MAE of the system \cite{han2019spin}. It has been found that $\sim$ 2 nm of C$_{60}$ exhibits magnetic moment $\sim$ 3$\mu_B$/cage at the epitaxial Fe/C$_{60}$ interface \cite{mallik2018effect}. There is a decrement in anisotropy in polycrystalline Fe/C$_{60}$ system whereas for polycrystalline Co/C$_{60}$ system anisotropy got enhanced. However, to the best of our knowledge no such basic study has been performed on CoFeB system. For spintronic application a low damping material is always desired as it directly affects the speed of a device.  The main advantage of taking CoFeB as a ferromagnet is that it exhibits low Gilbert damping parameter and it is amorphous in nature \cite{singh2019inverse}. It is very important to explore the effect of interface of such a system (CoFeB/OSC) to enrich our fundamental knowledge of spinterface. 

In this regard, we have prepared CoFeB/C$_{60}$ bilayer films and compared the magnetic properties to its reference CoFeB film. Also, we have varied the thickness of C$_{60}$ layer to qualitatively define the extent of spinterface and study the modifications in the basic magnetic properties. To study the qualitative nature of the interface, we have performed Kerr microscopy and ferromagnetic resonance (FMR) measurements.  

\section{Methods}

CoFeB reference film with 5 nm thickness and bilayer (CoFeB/C$_{60}$) samples have been deposited on Si (100) substrate in a multi-deposition high vacuum chamber manufactured by Mantis Deposition Ltd., UK. In the bilayer samples the thickness of CoFeB is fixed to 5 nm and the thickness of C$_{60}$ has been varied between 1.1 to 15 nm. The composition of CoFeB considered here is 40:40:20. The base pressure in the chamber was $5\times 10^{-8}$ mbar. CoFeB, C$_{60}$ and MgO layers have been deposited using DC sputtering, thermal evaporation  and e-beam evaporation techniques, respectively. The samples are named as S1, S2, S3, S4 and S5 for the thickness of C$_{60}$ ($t_{C_{60}}$) taken as 0, 1.1, 2, 5, 15 nm, respectively. The schematic of the sample structure is shown in Figure 1a (thicknesses not to scale). All the layers were deposited without breaking the vacuum to avoid oxidation and surface contamination. The deposition pressure was $5\times 10^{-3}$ mbar for CoFeB and $1 \times 10^{-7}$ mbar for C$_{60}$ and MgO evaporation. The deposition rates for CoFeB and C$_{60}$ layers were 0.1 and  $\sim$ 0.1 – 0.15 \AA/s, respectively. 2 nm of MgO has been deposited as a capping layer. C$_{60}$ layer has been deposited normal to the substrate whereas CoFeB plume was at 30$^{\circ}$ w.r.t the substrate normal due to chamber’s in-built geometry.

 To understand the growth of each layer and interfaces, cross-sectional TEM has been performed on sample S4 using a high-resolution transmission electron microscope (HRTEM) (JEOL F200, operating at 200 kV and equipped with a GATAN oneview CMOS camera). For the compositional analysis we have performed scanning transmission electron microscopy - energy dispersive X-ray spectroscopy (STEM - EDX). Selected area electron diffraction (SAED) has been performed on sample S4 to investigate the growth of the CoFeB and C$_{60}$ layers (supplementary information Figure S1).
X-ray reflectivity (XRR) has been performed on all the samples to know the exact thickness and roughness of all the layers (see Figure S2 and Table S1 in the supplementary information ).

We have measured the hysteresis loop and magnetic domain images at room temperature by magneto-optic Kerr effect (MOKE) based microscopy manufactured by Evico magnetics GmbH, Germany. Longitudinal hysteresis loops are recorded for $\pm$ 5 mT magnetic field by varying the angle ($\phi$) between the easy axis (EA) and the applied magnetic field direction. To measure the hysteresis loops along hard axis (HA), we have applied $\pm$ 17.5 mT magnetic field.

In order to determine the magnetic anisotropy constant and observe the anisotropy symmetry in the samples, angle dependent FMR measurements have been performed at a frequency of 7 GHz at 5$^{\circ}$ interval. During the measurement the sample was kept on the wide coplanar waveguide (CPW) in a flip-chip manner. An in-plane magnetic field (i.e, parallel to the sample plane) is applied to the sample (for the detailed measurement configuration, refer to the figure S4 in the supplementary information). Frequency dependent FMR measurements have been performed to calculate the Gilbert damping constant($\alpha$).

\section{Results and discussion}

	\begin{figure*}[h]
	\centering
	\includegraphics[width=1.0\linewidth]{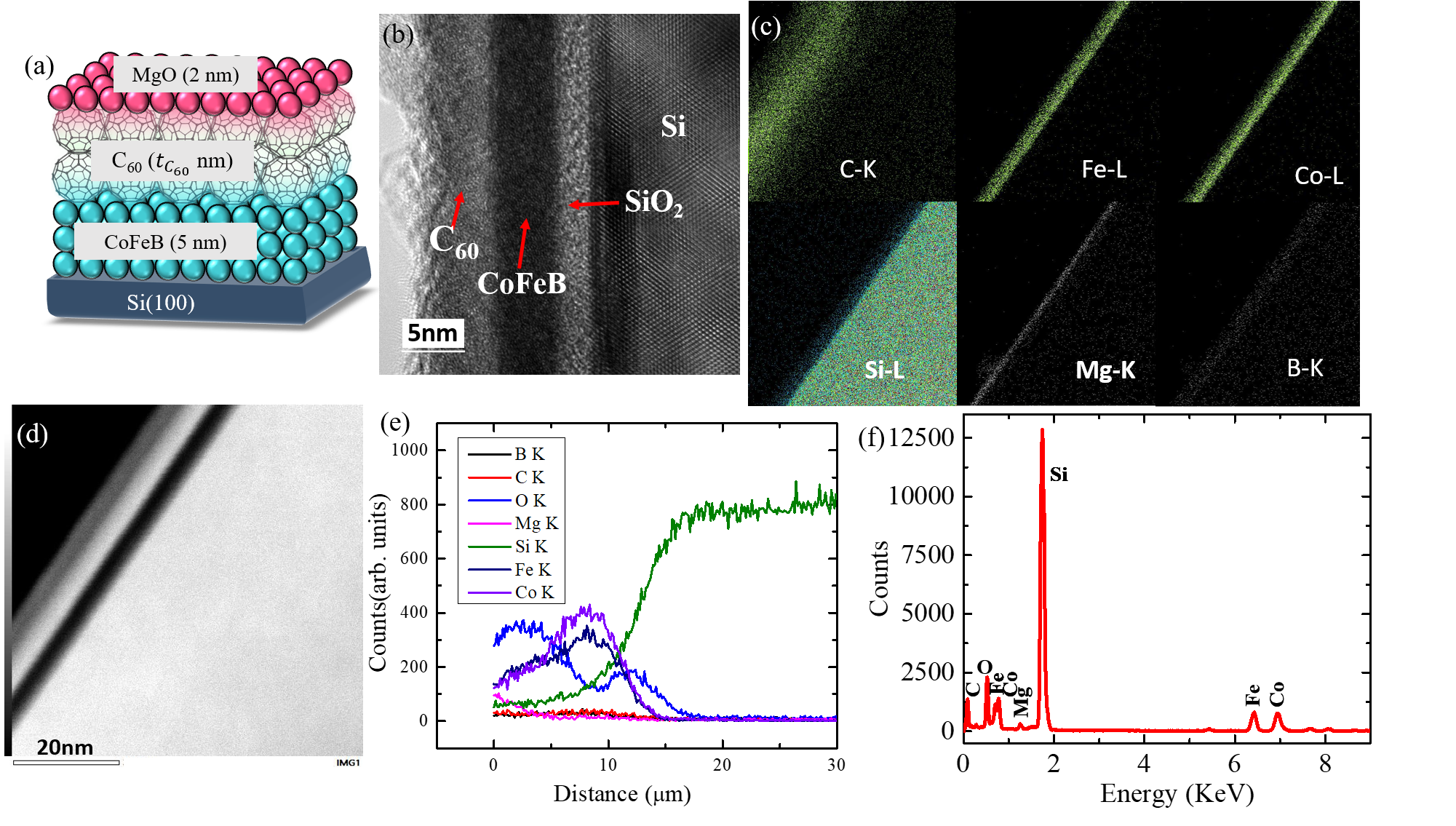}
	\caption{(a)Schematic of the sample structure. The thicknesses shown in this schematic is not to scale to the actual thicknesses of the samples. (b) Cross-sectional transmission electron microscopy (TEM) image of S4. (c) Elemental mapping for individual layers. (d)The region of the sample S4 where the STEM-EDX has been performed. (e) EDX line profile for each layer of the sample S4. (f) EDX spectrum of sample S4 showing the presence of different elements.}
	\label{fig:fig1}
\end{figure*}

High resolution TEM image is shown in Figure 1b and all the layers are marked separately. It shows the amorphous growth of CoFeB and $C_{60}$ (see supplementary information figure S1). Element specific mapping has been shown in Figure 1c. Figure 1d shows the STEM image, in which the brighter part indicates the layer of the element having high atomic number(Z). Presence of Boron(B) is not properly visible as it is a lighter atom. Figure 1e-f represent the EDX line profile and EDX spectra, respectively. The position of the Co and Fe peak at the same place indicates the formation of CoFeB alloy. EDX spectra shows the presence of C, Mg, O, Fe and Co elements in the sample.

\begin{figure*}[h]
	\centering
	\includegraphics[width=1.0\linewidth]{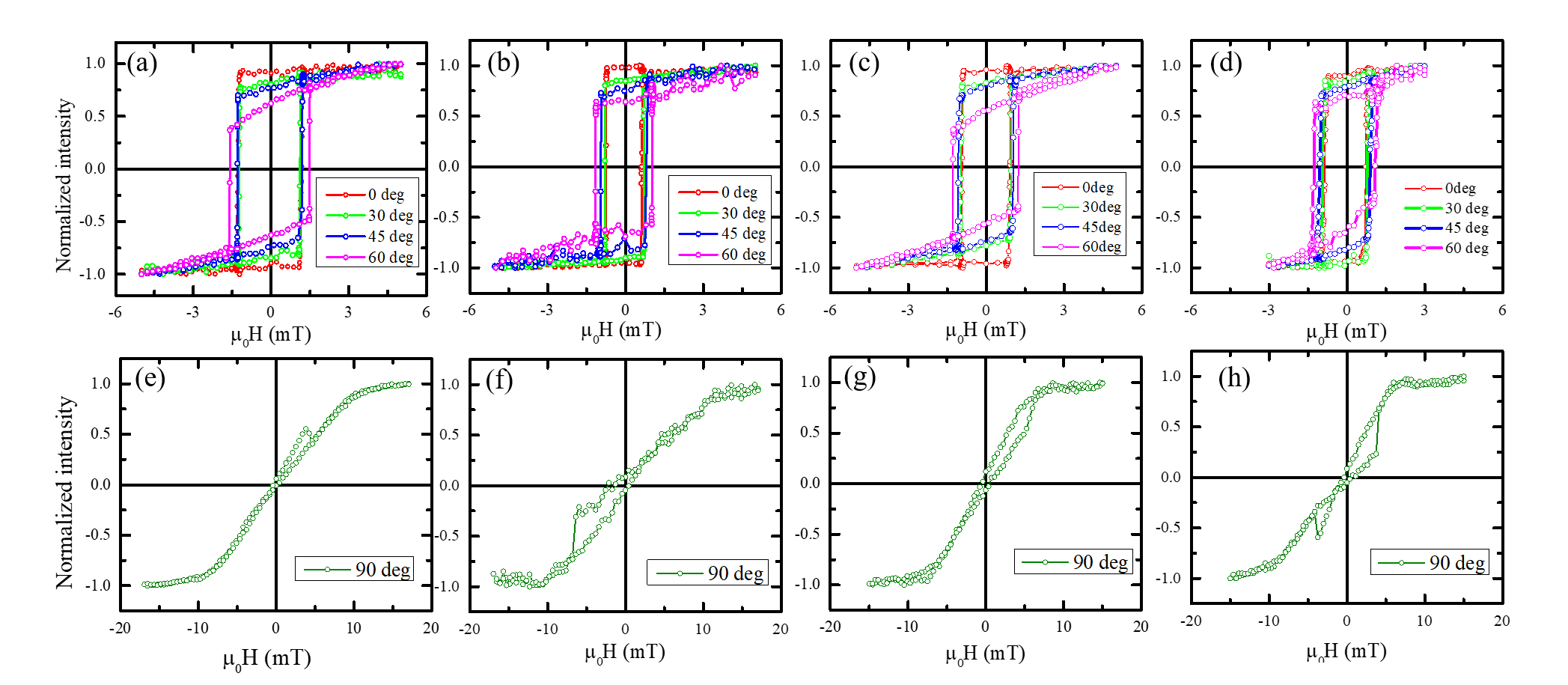}
	\caption{Hysteresis loops measured by magneto optic Kerr effect (MOKE)  microscopy at room temperature in longitudinal mode by varying the angle ($\phi$) between the EA and the applied magnetic field direction for the samples (a) S1, (b) S2, (c) S4 and (d) S5. (e) - (h) represent the hysteresis loops measured along 90$^{\circ}$ w.r.t EA for S1, S2, S4 and S5, respectively.} 
	\label{fig:fig2}
\end{figure*}

\begin{figure*}[h]
	\centering
	\includegraphics[width=1\linewidth]{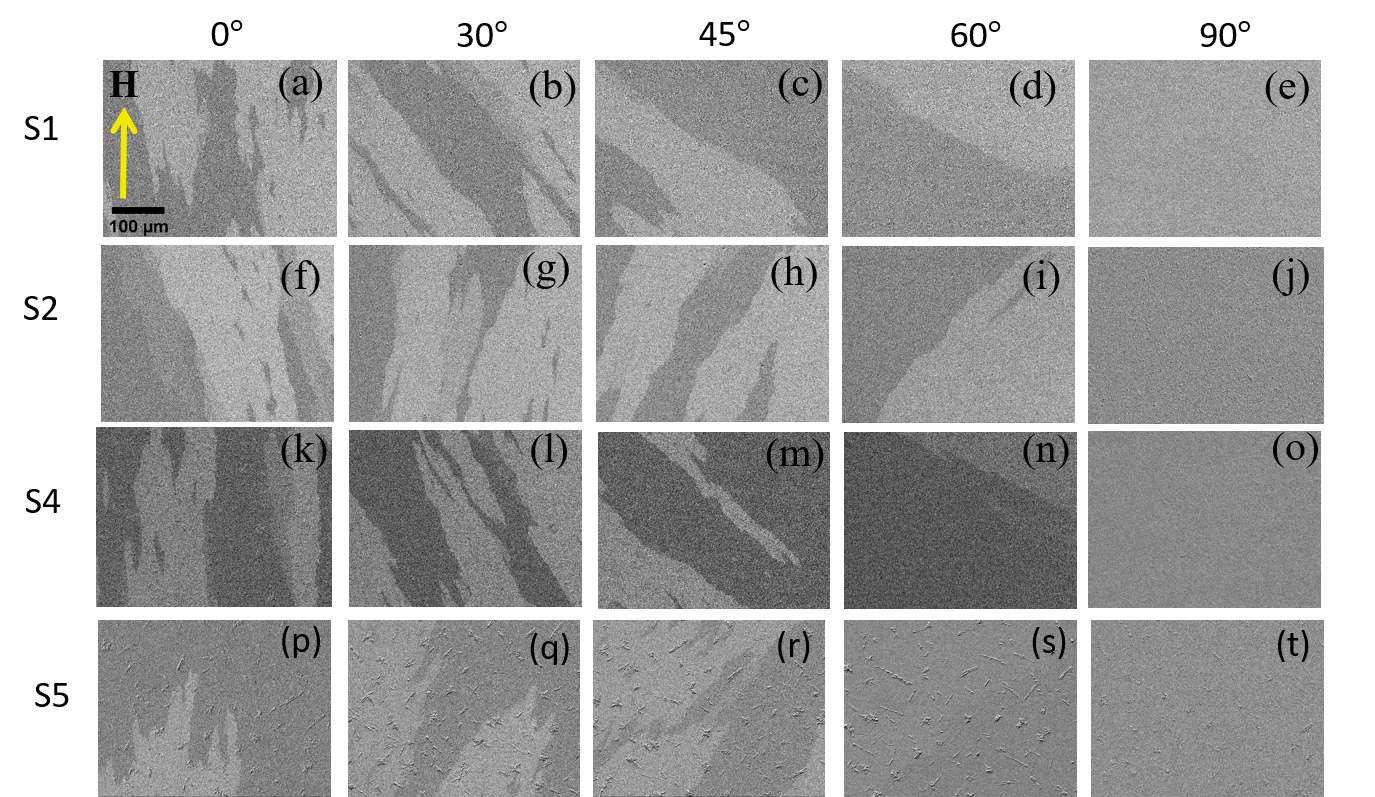}
	\caption{Domain images near $H_{C}$ for samples S1, S2, S4 and S5 are shown in (a) - (e), (f) - (j), (k) - (o) and (p) – (t), respectively. The scale bars of the domain images for all the samples are same and shown in image (a). The applied field ($H$) direction shown in image (a) was kept constant for all the measurements and the sample was only rotated to capture the domain images at different $\phi$.}
	\label{fig:fig3}
\end{figure*}

Figure 2a-d show the in-plane angle ($\phi$) dependent hysteresis loops measured using longitudinal magneto optic Kerr effect (LMOKE) microscopy at room temperature for the samples S1, S2, S4 and S5, respectively. $\phi$ is defined as the angle between the EA and the applied magnetic field direction. The hysteresis loops along 90$^{\circ}$ w.r.t. EA are shown in Figure 2e-h for the samples S1, S2, S4 and S5, respectively. Angle dependent hysteresis loops show that the magnetic HA of the samples is at 90$^{\circ}$ w.r.t the EA, which marks the presence of uniaxial anisotropy in the system. The easy axis of the anisotropy lies in-plane at an angle of 90$^{\circ}$ w.r.t the projection of the plume direction. The CoFeB target is at an angle of 30$^{\circ}$ w.r.t. the substrate normal due to the in-built geometry of the deposition system. Such kind of oblique angle deposition induced uniaxial anisotropy has been reported earlier\cite{smith1960oblique,bubendorff2006origin,chowdhury2016study,mallick2018tuning,mallik2019enhanced,mallik2018study,mallik2017effect,nayak2019tuning}. It should be noticed that there is a change in coercive field ($H_{C}$) in the bilayer samples as compared to the single layer reference sample. The values of $H_{C}$ are 1.23, 0.71, 0.91 and 0.81 mT for the samples S1, S2, S4 and S5, respectively. The $H_{C}$ for the bilayer samples S2 to S5 are comparable. However, the decrease in $H_{C}$ from single layer CoFeB to bilayers CoFeB/C$_{60}$ can be attributed to the formation of a spinterface between the CoFeB and C$_{60}$ interface. In our previous study we have shown that the orbital hybridization at the FM (Fe or Co)/C$_{60}$ interface promote the change in anisotropy of the system \cite{mallik2018effect,mallik2019tuning,mallik2019enhanced}.

The square shaped loop along EA indicates the magnetization reversal is happening via domain wall motion whereas, along HA the reversal occurs via coherent rotation. The magnetization reversal is studied as a function of $\phi$. By varying the angle ($\phi$) w.r.t easy axis (0$^{\circ}$), we have recorded the domain images near the $H_{C}$ at $\phi$ = 0$^{\circ}$, 30$^{\circ}$, 45$^{\circ}$, 60$^{\circ}$ and 90$^{\circ}$.  Figure 3a-e, Figure 3f-j, Figure 3k-o and Figure 3p-t show the magnetic domain images near $H_{C}$ for the samples S1, S2, S4 and S5, respectively. Branched domains have been observed in all the samples due to the amorphous growth of CoFeB. Domain images captured at different applied magnetic fields along EA for samples S1, S2, S4 and S5, are shown in Figure S3 in supplementary information. For the samples S2 and S5 the tilt of the domains are opposite to S1 and S4. This opposite tilt is due to the anti-clockwise (S2 and S5) and clockwise (S1 and S4) rotation of the sample stage w.r.t EA during measurement. It is noted that the $H_{C}$ is different between the reference CoFeB and the bilayer samples. However, the change in domain structure is not significant between the single layer CoFeB and the bilayer CoFeB/C$_{60}$ samples. In our earlier reports it has been shown that the change in domain shape and size is significant in other ferromagnetic/OSC systems such as Co/C$_{60}$, Fe/C$_{60}$ \cite{mallik2018effect,mallik2019tuning,mallik2019enhanced}. In epitaxial Fe/C$_{60}$ system the magnetization reversal process was different between the reference Fe and the Fe/C$_{60}$ bilayer systems \cite{mallik2018effect}. In case of polycrystalline Fe/C$_{60}$ system the domain size got reduced for the bilyaers as compared to the reference Fe film \cite{mallik2019tuning}. However, in case of polycrystalline Co/C$_{60}$, the domain size increased for the bilayers as compared to the reference Co film \cite{mallik2019enhanced}. In this study we considered CoFeB system and the domain shape and size are comparable between the bilayers and the reference sample. The origin to this may be investigated theoretically in future work.

\begin{figure*}[h]
	\centering
	\includegraphics[width=1.0\linewidth]{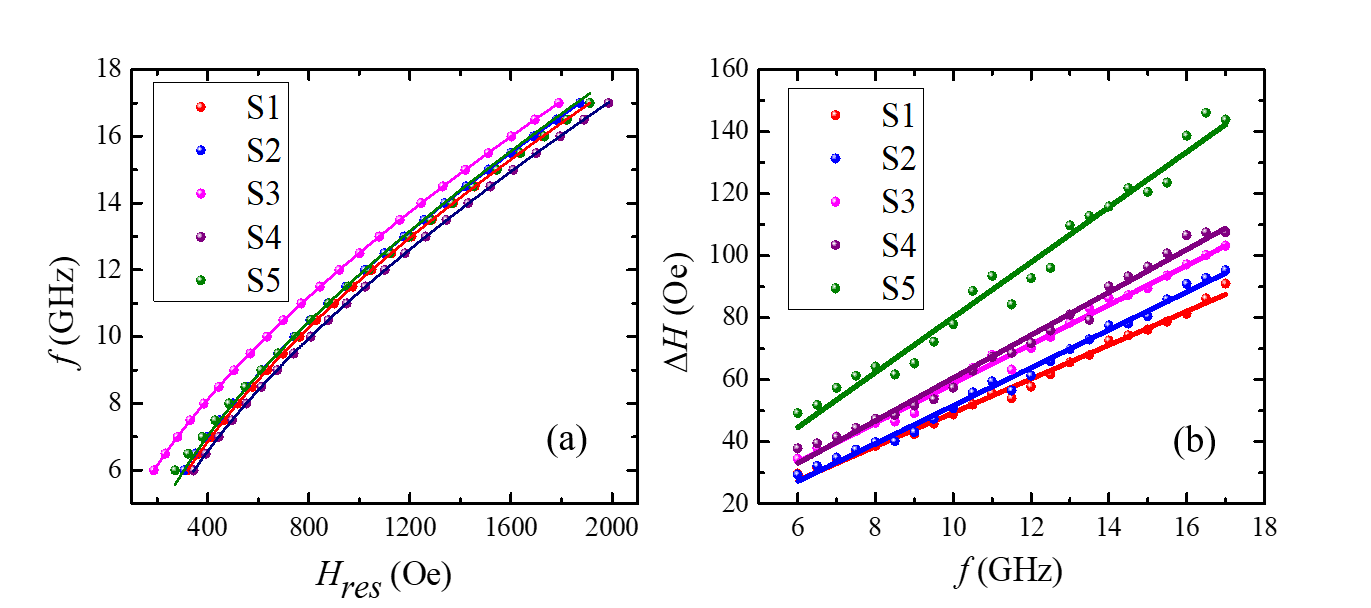}
	\caption{ $f$ vs $H_{res}$ and $\Delta H$ vs $f$ plots for S1, S2, S3, S4 and S5 are shown in (a) and (b), respectively. Solid circles represent the experimental data, while the solid lines are the best fits to the eqs. 2 and 3.}
	\label{fig:fig4}
\end{figure*}

We have further invesigated the magnetization dynamics by performing the frequency dependent FMR measurement. The experimental data has been fitted using a Lorentzian function (eq. 1), where $\Delta H$, $H_{res}$, A$_{1}$ and A$_{2}$ are linewidth, resonance field, anti-symmetric and symmetric components, respectively \cite{singh2017study}.

\begin{eqnarray}
	FMR_{signal} = A_{1} \frac{4\Delta H (H-H_{res})}{(4(H-H_{res}))^2+(\Delta H)^2} - A_{2} \frac{(\Delta H)^2- 4(H-H_{res})^2}{(4(H-H_{res}))^2+(\Delta H)^2}+offset 	
\end{eqnarray}


The plots of $f$ vs $H_{res}$ and $\Delta H$ vs $f$ are shown in Figure 4a and Figure 4b, respectively. The effective damping constant ($\alpha$) has been determined by fitting the eqs. 2 and 3 \cite{singh2017study,kittel1948theory,heinrich1985fmr}:

\begin{eqnarray}
	f= \frac{\gamma}{2\pi}\sqrt{(H_{K}+H_{res})(H_{K}+H_{res}+4\pi M_{eff})}
\end{eqnarray}

where, $\gamma$ (gyromagnetic ratio) = $g \mu_{B}/\hbar$
and $g$, $\mu_{B}$, $\hbar$, $H_{K}$ are Lande-g factor, Bohr magneton, reduced Planck's constant and anisotropy field, respectively.

\begin{eqnarray}
	\Delta H= \Delta H_{0} + \frac{4 \pi \alpha f}{\gamma}
\end{eqnarray}

where, $\Delta H_{0}$ is the inhomogeneous line width broadening which depends on the magnetic inhomogeneity of the sample. $\alpha$ values for the samples S1, S2, S3, S4 and S5 are 0.0095$\pm$0.0002, 0.0106$\pm$0.0002, 0.0110$\pm$0.0002, 0.0124$\pm$0.0003 and 0.0169$\pm$0.0006 , respectively. It has been observed that $\alpha$ increases after introducing a C$_{60}$ layer and it further increased with increasing the C$_{60}$ thickness. This increase in $\alpha$ might be due to the interface roughness or other effects such as spin pumping at the interface{\cite{sharangi2021spin}}.

\begin{figure}[h]
	\centering
	\includegraphics[width=0.75\linewidth]{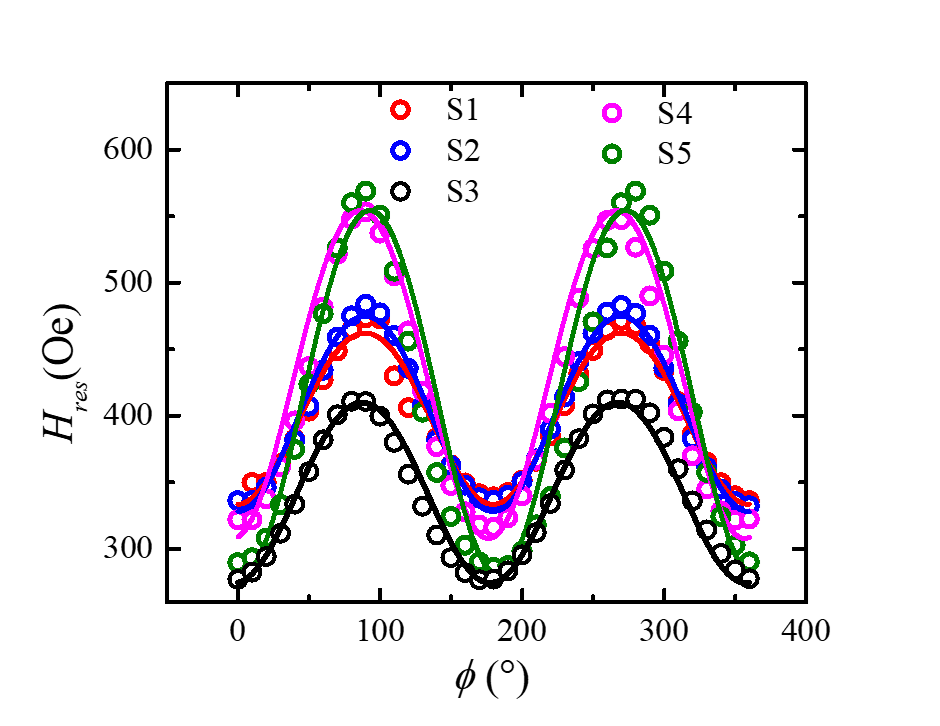}
	\caption{Angle dependent resonance field ($H_{res}$) plot for all the five samples to calculate the anisotropy constants of the system. The measurement was performed at room temperature and a fixed frequency of 7 GHz. Open circles represent the experimental data, while the solid lines are the best fits.}
	\label{fig:fig5}
\end{figure}

To quantify the change in anisotropy in all the samples we have performed in-plane angle dependent FMR measurements at a fixed frequency of 7 GHz. Resonance field ($H_{res}$) has been measured by rotating the sample w.r.t the applied magnetic field in 5$^{\circ}$ intervals. 

$H_{res}$ vs $\phi$ plots have been shown in Figure 5 to calculate the anisotropy constants of the system. The open circles represent the raw data and the solid lines are the best fits. The experimental data is fitted using Landau-Lifshitz-Gilbert (LLG) equation \cite{pan2017role}: 

	\begin{equation}
	f = \frac{\gamma}{2\pi} ((H+\frac{2K_{2}}{M_{S}}Cos2\phi)(H+4\pi M_{S}+ \frac{2K_{2}}{M_{S}}Cos^{2}\phi))^{1/2}
\end{equation}

where, $K_{2}$ is the in-plane uniaxial anisotropy constant, $\phi$ is the in-plane angle between the easy axis w.r.t the applied magetic field direction and $M_{S}$ is the saturation magnetization.

\begin{table}[]
	\centering
	\caption { The value of $K_{2}$ for all the samples extracted from the fitting of LLG equation.}
	
	\begin{tabular}{|l|l|}
		\hline
		\multicolumn{1}{|c|}{Sample} & \multicolumn{1}{c|}{$K_{2}$ (erg/cc)} \\ \hline
		S1                           & 2.4$\times$$10^{4}$        \\ \hline
		S2                           & 2.9$\times$$10^{4}$         \\ \hline
		S3                           & 3.1$\times$$10^{4}$         \\ \hline
		S4                           & 4.1$\times$$10^{4}$         \\ \hline
		S5                           & 4.3$\times$$10^{4}$          \\ \hline
	\end{tabular}
\end{table}

The $K_{2}$ values extracted from the fitting are listed in Table 1. It has been observed that by introducing a $C_{60}$ layer the anisotropy of the system increased. The possible reason behind the enhancement in the magnetic anisotropy is the formation of spinterface at the CoFeB/$C_{60}$ interface. The anisotropy increases from 2.9 $\times 10^{4} $ to 3.1 $\times 10^{4}$ erg/cc when the $C_{60}$ thickness is varied from 1.1 to 2 nm. With further increase in $C_{60}$ thickness (at 5 nm), the anisotropy become 4.1 $\times 10^{4}$ erg/cc.  There is a small change in the anisotropy ($4.1 \times 10^{4}$ to $4.3 \times 10^{4}$ erg/cc) when $C_{60}$ thickness increases from 5 to 15 nm. After a certain thickness of $C_{60}$ layer, the spinterface thickness remains almost constant. However, the exact thickness of the spinterface for amorphous CoFeB/$C_{60}$ system is not known. In future polarized neutron reflectivity (PNR) experiment may be carried out to evaluate the spinterface thickness.

\section{Conclusion}
We have studied the effect of C$_{60}$ on the magnetization reversal and the magnetic anisotropy of a low damping amorphous CoFeB layer. In comparison to the single layer CoFeB sample the magnetic anisotropy constant has been increased for the CoFeB/C$_{60}$ bilayer samples. Further from the Kerr microscopy measurements it is observed that there is a negligible change in the branch domain pattern in the samples. The enhancement in magnetic anisotropy might be the result of $d-p$ hybridization between the CoFeB and C$_{60}$ layer. This study reveals that one can enhance the anisotropy of a ferromagnetic CoFeB system by introducing a C$_{60}$ layer, which can be suitable for future spintronics devices. Further in future, the nature of spinterface such as thickness, magnetic moment per atom etc. should be investigated by experimental methods such as polarized neutron reflectometry. The results presented here might bring interest to study similar system theoretically to elucidate the exact nature of spinterface and the origin behind it.

\section{Supporting Information}
	
	Selected area electron diffraction (SAED) on sample S4 is shown in Figure S1. XRR data with the best fits for samples S1 to S5 are shown in Figure S2. MOKE hysteresis loops with corresponding domain images along the EA for samples S1, S2, S4 and S5 are shown in Figure S3. The schematic of the FMR measurement set-up and applied field direction is shown in Figure S4.

\begin{acknowledgement}

We sincerely thank Dr. Tapas Gosh and Mr. Pushpendra Gupta for helping in TEM imaging. The authors also want to thank Dr. Ashutosh Rath for valuable discussion regarding the SAED images. The authors also acknowledge Department of Atomic Energy, and Department of Science and Technology - Science and Engineering Research Board, Govt. of India (DST/EMR/2016/007725) for the financial support.

\end{acknowledgement}


\bibliography{reference}


	

\end{document}






\section{Structural information}
In order to investigate the growth of deposited layers we have performed selected area electron diffraction (SAED) on sample S4 (i.e., CoFeB(5nm)/C$_{60}$ (5 nm)/MgO(2nm)). The SAED image (Figure S1) shows the diffuse rings, which confirms the amorphous growth of CoFeB and C$_{60}$ layers. Further, to know the structural information (thikness, roughness)$,$ we have performed X-ray reflectivity (XRR) measurements on all the samples. We have fitted the data by using GenX software.  Figure S2a-e show the XRR data and best fits for samples S1, S2, S3, S4 and S5$,$ respectively. The extracted thickness ($t$) and roughness ($\sigma$) of the layers are shown in Table S1.

\begin{figure*}
	\centering
	\includegraphics[width=0.5\linewidth]{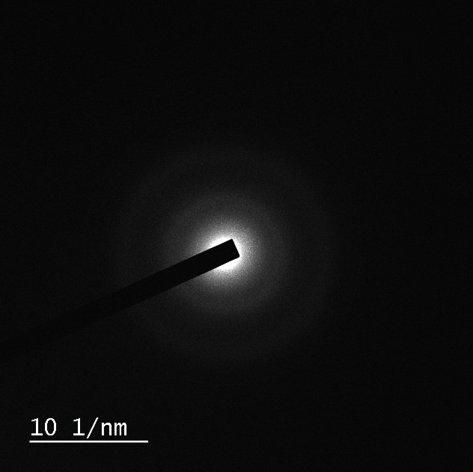}
	\caption{SAED image of sample S4.} \label{fig:fig-S1}
\end{figure*}

\begin{figure*}
	\centering
	\includegraphics[width=0.9\linewidth]{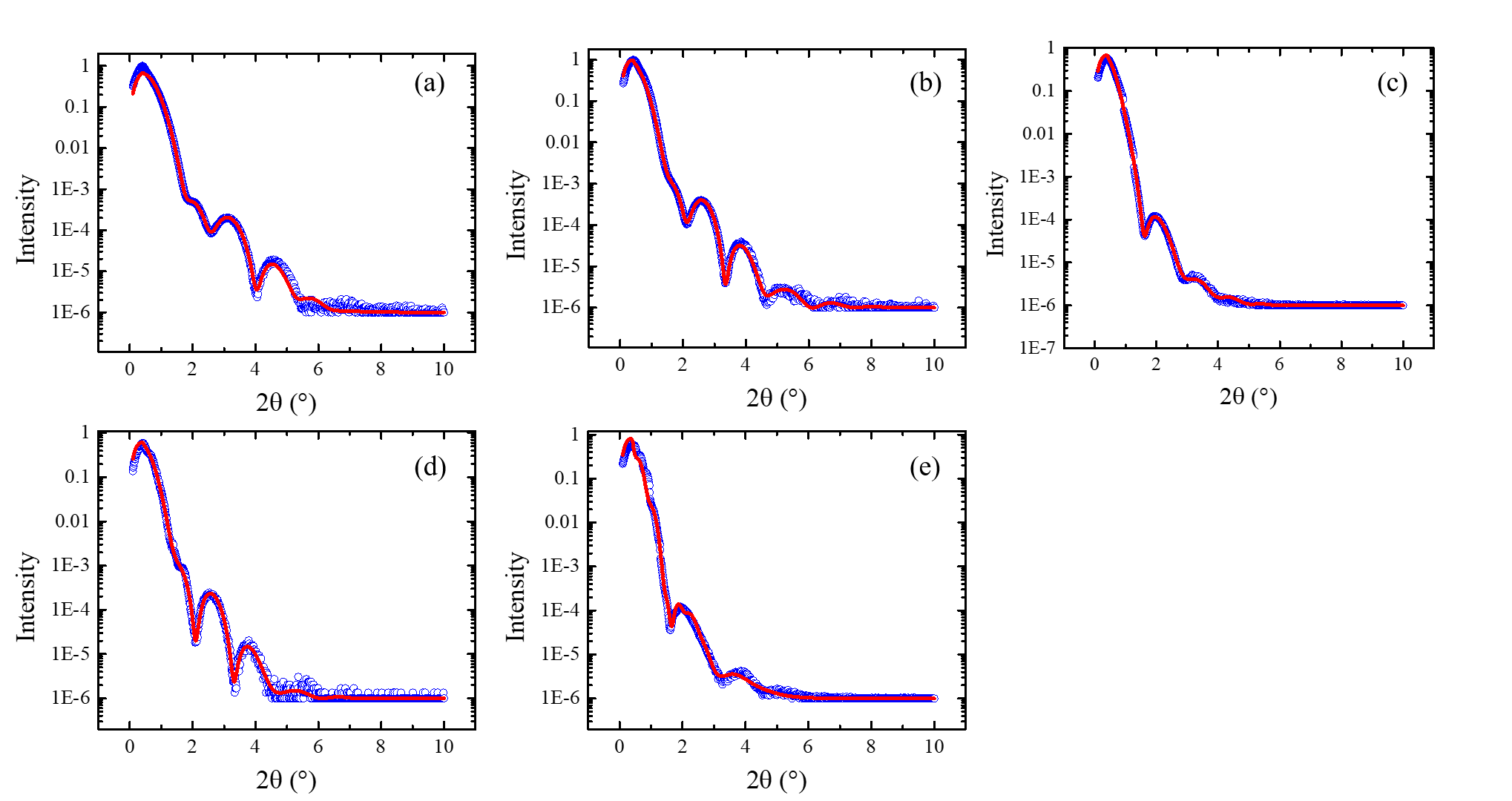}
	\caption{XRR data and the best fits for samples S1, S2, S3, S4 and S5 are shown in (a), (b)$,$ (c), (d) and (e), respectively. The blue open circles are experimental data and the red solid lines represent the best fit using GenX software. The parameters extracted from the best fits are shown in Table S1.}	\label{figure S2}
\end{figure*}

\begin{table*}
	
	\caption {Parameters obtained from XRR fits }
	\begin{tabular}{|c|c|c|c|c|c|c|c|c|c|c|}
		\hline
		& \multicolumn{2}{c|}{Sample S1}   & \multicolumn{2}{c|}{Sample S2}& \multicolumn{2}{c|}{Sample S3} & \multicolumn{2}{c|}{Sample S4} &\multicolumn{2}{c|}{Sample S5} \\ \hline
		Layers & $t$(nm) & $\sigma$(nm) & $t$(nm) & $\sigma$(nm) & $t$(nm) & $\sigma$(nm) & $t$(nm) & $\sigma$(nm)& $t$(nm) & $\sigma$(nm) \\ \hline
		CoFeB    & 5.50      & 0.91      & 5.60    & 0.93  & 5.50    & 0.87 & 5.00    & 0.81  & 5.45    & 0.96 \\ \hline
		C$_{60}$     & -   & -   & 1.10     & 0.20  & 2.00    & 0.60  & 5.00   & 0.52  & 15.00   & 1.80  \\ \hline
		MgO     & 1.95       & 0.61     & 1.90    & 0.57 & 1.80    & 0.51  & 2.09    & 0.76  & 1.98   & 0.75 \\ \hline	
	\end{tabular}
	\label{Table S1}
\end{table*}

\begin{figure*}
	\centering
	\includegraphics[width=1.0\linewidth]{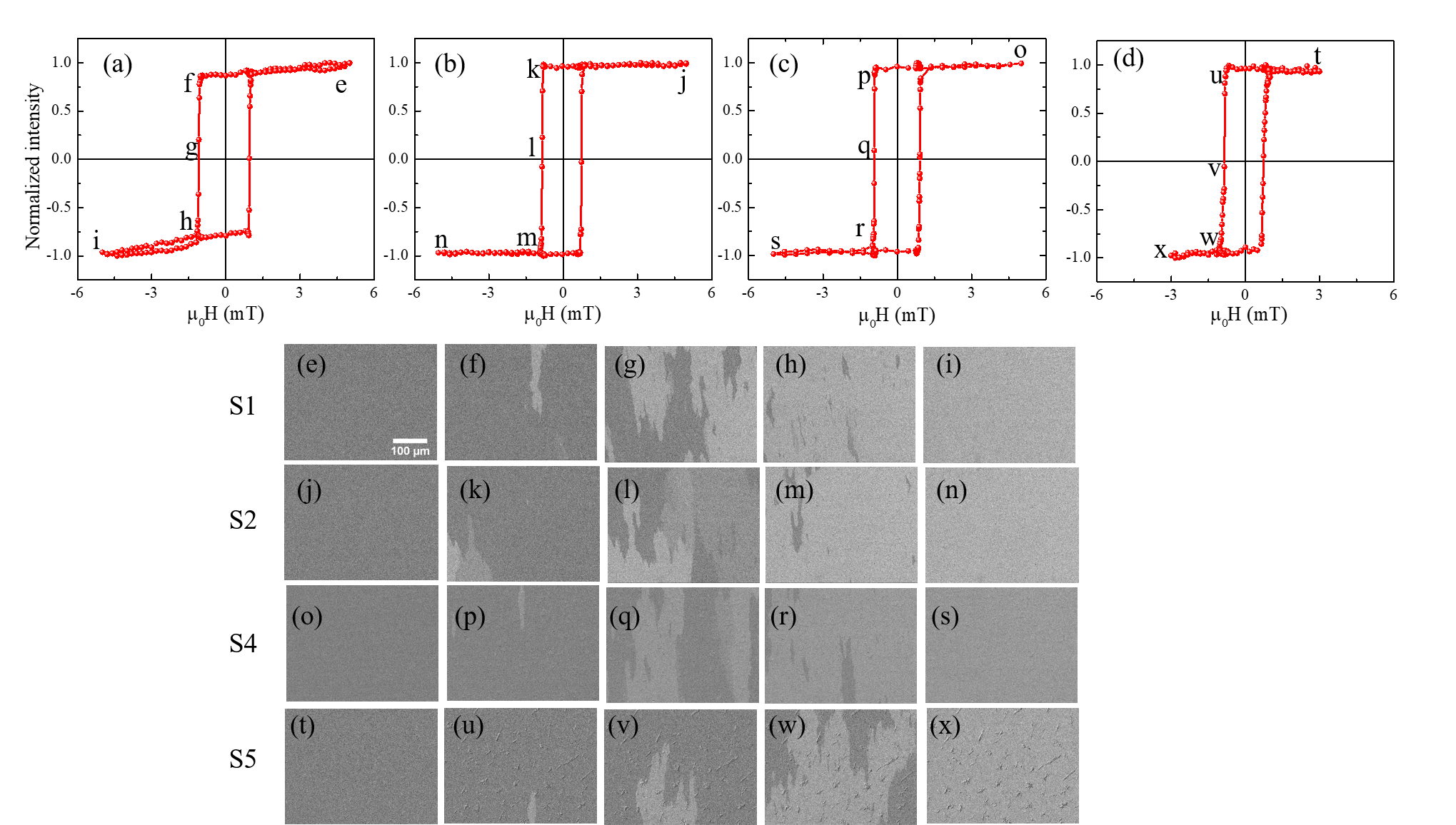}
	\caption{(a), (b), (c) and (d) show the hysteresis loops for the samples S1, S2, S4 and S5, respectively, along the easy axis (EA). The domain images at different applied fields (marked e to x in hysteresis loops) for samples S1, S2, S4 and S5 are shown in (e-i), (j-n), (o-s) and (t-x), respectively. The scale bars of the domain images for all the samples are same and shown in image (e).}
	\label{figS3}
\end{figure*}

\section{Hysteresis loop and domain imaging}

Figure S3a-d show the hysteresis loops for the samples S1, S2, S4 and S5, respectively, along the easy axis (EA). The domain images are shown in Figure S3e-x are also marked in the hysteresis loops. Figure S3e, Figure S3j, Figure S3o and Figure S3t represent the domain images for S1, S2, S4, S5 captured at positive saturation field. Similarly, Figure S3f, Figure S3k, Figure S3p, Figure S3u are the domain images near nucleation. Figure S3g, Figure S3l, Figure S3q, Figure S3v show the domain images which are captured near coercive field and Figure S3h, Figure S3m, Figure S3r, Figure S3w are captured near negative saturation for samples S1, S2, S4 and S5, respectively. Further, domain images captured at negative saturation field are shown in Figure S3i, Figure S3n, Figure S3s, Figure S3x for S1, S2, S4 and S5, respectively.

During FMR measurement we applied an in-plane magnetic field (i.e, parallel to the sample plane). Figure S4 shows the schematic of the FMR measurement set-up and the applied field direction.

\begin{figure*}
	\centering
	\includegraphics[width=0.5\linewidth]{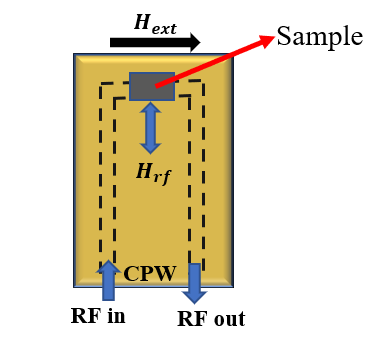}
	\caption{Schematic representation of FMR measurement set-up. Sample is placed on the cpw in a flip chip manner. The applied field ($H_{ext}$) is parallel to the sample plane and perpendicular to the rf field ($H_{rf}$).}
	\label{figS4}
\end{figure*}
